\documentstyle[11pt,amssymb,epsf]{article}

\makeatletter
\@addtoreset{equation}{section}
\makeatother

\def\ben{\begin{equation}}
\def\een{\end{equation}}

  \let\n=\nu

\let\C=\Chi

\def\nn{\nonumber} \def\bd{\begin{document}} \def\ed{\end{document}}
\def\ds{\documentstyle} \let\fr=\frac \let\bl=\bigl \let\br=\bigr
\let\Br=\Bigr \let\Bl=\Bigl
\let\bm=\bibitem
\let\na=\nabla
\let\pa=\partial \let\ov=\overline
\newcommand{\be}{\begin{equation}}
\newcommand{\ee}{\end{equation}}
\def\ba{\begin{array}}
\def\ea{\end{array}}
\def\ft#1#2{{\textstyle{{\scriptstyle #1}\over {\scriptstyle #2}}}}
\def\fft#1#2{{#1 \over #2}}
\def\del{\partial}
\def\vp{\varphi}
\def\sst#1{{\scriptscriptstyle #1}}
\def\oneone{\rlap 1\mkern4mu{\rm l}}
\def\td{\tilde}
\def\wtd{\widetilde}
\def\ie{\rm i.e.\ }
\def\dalemb#1#2{{\vbox{\hrule height .#2pt
        \hbox{\vrule width.#2pt height#1pt \kern#1pt
                \vrule width.#2pt}
        \hrule height.#2pt}}}
\def\square{\mathord{\dalemb{6.8}{7}\hbox{\hskip1pt}}}
\newcommand{\ho}[1]{$\, ^{#1}$}
\newcommand{\hoch}[1]{$\, ^{#1}$}
\newcommand{\bea}{\begin{eqnarray}}
\newcommand{\eea}{\end{eqnarray}}
\newcommand{\ra}{\rightarrow}
\newcommand{\lra}{\longrightarrow}
\newcommand{\Lra}{\Leftrightarrow}
\newcommand{\ap}{\alpha^\prime}
\newcommand{\bp}{\tilde \beta^\prime}
\newcommand{\tr}{{\rm tr} }
\newcommand{\Tr}{{\rm Tr} }
\def\0{{\sst{(0)}}}
\def\1{{\sst{(1)}}}
\def\2{{\sst{(2)}}}
\def\3{{\sst{(3)}}}
\def\4{{\sst{(4)}}}
\def\5{{\sst{(5)}}}
\def\6{{\sst{(6)}}}
\def\7{{\sst{(7)}}}
\def\8{{\sst{(8)}}}
\def\n{{\sst{(n)}}}
\def\cA{{{\cal A}}}
\def\cB{{{\cal B}}}
\def\cF{{{\cal F}}}
\def\cH{{{\cal H}}}
\def\tV{\widetilde V}
\def\tW{\widetilde W}
\def\tH{\widetilde H}
\def\tE{\widetilde E}
\def\tF{\widetilde F}
\def\tA{\widetilde A}
\def\im{{i}}
\def\tY{{{\wtd Y}}}
\def\ep{{\epsilon}}
\def\vep{{\varepsilon}}
\def\R{\rlap{\rm I}\mkern3mu{\rm R}}
\def\bD{{{\bar D}}}

\def\R{\rlap{\rm I}\mkern3mu{\rm R}}
\def\bD{{{\bar D}}}
\def\R{{{\Bbb R}}}
\def\C{{{\Bbb C}}}
\def\H{{{\Bbb H}}}
\def\CP{{{\Bbb C}{\Bbb P}}}
\def\RP{{{\Bbb R}{\Bbb P}}}
\def\Z{{{\Bbb Z}}}
\def\bA{{{\Bbb A}}}
\def\bB{{{\Bbb B}}}
\def\bC{{{\Bbb C}}}
\def\bD{{{\Bbb D}}}
\def\bE{{{\Bbb E}}}
\def\bZ{{{\Bbb Z}}}
\def\Re{{{\frak{Re}}}}
\def\Im{{{\frak{Im}}}}
\def\cosec{{\,\hbox{cosec}\,}}
\def\Gm{{\Gamma_{\!\! -}}}
\def\Gp{{\Gamma_{\!\! +}}}
\def\stan{{standard }}
\def\nonstan{{supernumerary }}

\newcommand{\tamphys}{\it Center for Theoretical Physics,
Texas A\&M University, College Station, TX 77843}

\newcommand{\upenn}{\it Department of Physics and Astronomy,\\ University
of Pennsylvania, Philadelphia, PA 19104}

\newcommand{\brussels}{\it Physique Th\'eorique et Math\'ematique,
Universit\'e Libre de Bruxelles,\\ Campus Plaine C.P. 231, B-1050
Bruxelles, Belgium}

\newcommand{\auth}{J. Kerimo\hoch{\dagger}, James T. Liu\hoch{*1}, 
H. L\"u\hoch{\dagger2} and C.N. Pope\hoch{\dagger2}}

\begin{document}
\begin{flushright}

MIFP-04-01\ \ \ MCTP-04-01\\
{\bf hep-th/0401001}\\
January\  2004
\end{flushright}

\vspace{10pt}

\begin{center}

{\large {\bf Variant ${\cal N}=(1,1)$ Supergravity and 
(Minkowski)$_4\times S^2$
Vacua}}

\vspace{20pt}
\auth

\vspace{20pt}

{\it \hoch{\dagger} George P. and Cynthia W. Mitchell
Institute for Fundamental Physics,\\ Texas A\& M University,
College Station, TX 77843-4242, USA}

\vspace{20pt}

{\it \hoch{*} Michigan Center for Theoretical Physics\\
University of Michigan, Ann Arbor, MI 48109-1120}


\vspace{40pt}

\underline{ABSTRACT}
\end{center}

    We construct the fermionic sector and supersymmetry transformation
rules of a variant ${\cal N}=(1,1)$ supergravity theory obtained by
generalized Kaluza-Klein reduction from seven dimensions.  We show
that this model admits both (Minkowski)$_4\times S^2$ and
(Minkowski)$_3\times S^3$ supersymmetric vacua.  We perform a
consistent Kaluza-Klein reduction on $S^2$ and obtain $D=4$, ${\cal
N}=2$ supergravity coupled to a vector multiplet, which can be
consistently truncated to give rise to $D=4$, ${\cal N}=1$
supergravity with a chiral multiplet.

{\vfill\leftline{}\vfill \vskip 10pt \footnoterule {\footnotesize
\hoch{1} Research supported in part by DOE grant
DE-FG02-95ER40899.}\vskip 2pt
{\footnotesize
\hoch{2} Research supported in part by DOE grant
DE-FG03-95ER40917.}

}

\pagebreak
\setcounter{page}{1}

\newpage

\section{Introduction}

The notion of higher dimensional theories and their Kaluza-Klein
compactifications has played a central r\^ole in many modern approaches
to unification, and especially M-theory.  In such models, it is essential
on phenomenological grounds that the resulting lower dimensional theory
yields a realistic spectrum of chiral fermions as well as correct strength
couplings and a small (or vanishing) cosmological constant.  Thus
while spheres appear to be natural and simple candidates for an internal
space, they are often rejected as they would typically give rise to
a lower dimensional theory with a large cosmological constant (directly
related to the curvature of the sphere).

However, a rare example of a sphere reduction admitting a flat Minkowski
spacetime has been known for a while.  This is the Salam-Sezgin model
\cite{salamsezgin}, which is essentially gauged ${\cal N}=(1,0)$
supergravity in six dimensions admitting a (Minkowski)$_4\times S^2$
vacuum.  This model has the added feature that the vanishing of the
cosmological constant in four dimensions typically arises naturally
based on an interplay between the six-dimensional potential and the
$U(1)$ monopole flux on the $S^2$.  This self-tuning mechanism has
been highlighted recently in \cite{quevedo1,quevedo2} where it was
noted that the four-dimensional cosmological constant is protected
against large contributions in a Salam-Sezgin braneworld scenario,
even after supersymmetry breaking on the branes.  (Note, however, that
this self-tuning presupposes the existence of a (Minkowski)$_4$ vacuum
to start with \cite{guven}.)  The $S^2$ reduction of the Salam-Sezgin model
was recently examined in \cite{gibbonpopes2}, and its lower dimensional
spectrum was analyzed.

Recently a variant ${\cal N}=(1,1)$ supergravity in $D=6$ was obtained
from generalized Kaluza-Klein reduction; as in the Salam-Sezgin model,
this theory also admits a (Minkowski)$_4\times S^2$ vacuum solution
\cite{kerimo}.  Unlike the Salam-Sezgin model, however, the variant ${\cal
N}=(1,1)$ supergravity includes vector fields in the gravity multiplet
itself.  In this case, the turning on of a six-dimensional potential is
related to mass generation for one of the graviphotons.  The bosonic
equations of motion for this model were obtained in \cite{kerimo},
where it was further argued that the model is indeed supersymmetric.
In this paper, we complete the reduction of \cite{kerimo} and prove its
supersymmetry by obtaining a complete set of supersymmetry transformation
rules via Kaluza-Klein reduction of the original $D=7$ theory.

After obtaining the complete supersymmetry transformations of
the variant ${\cal N}=(1,1)$ theory, we investigate some of its
spontaneous compactifications.  As in the ${\cal N}=(1,0)$ model, we
find that it can also be consistently reduced on a 2-sphere to give
rise to four-dimensional ${\cal N}=2$ supergravity coupled to a single
vector multiplet.  This can be further truncated to yield ${\cal N}=1$
supergravity coupled to a chiral multiplet.  We further demonstrate that,
in contrast to the ${\cal N} =(1,0)$ theory, this model admits also a
supersymmetric (Minkowski)$_3\times S^3$ vacuum.

We begin in section 2 by investigating the fermionic sector of the $D=7$
to $D=6$ reduction.  After constructing the fermionic reduction ansatz,
we provide the complete supersymmetry transformations of the variant
${\cal N} =(1,1)$ supergravity.  In section 3, we examine the reduction
to (Minkowski)$_4\times S^2$.  Since the resulting theory is simply
${\cal N} =2$ supergravity coupled to a vector, it naturally admits
supersymmetric dyonic black hole solutions.  The black hole and its
lifting back to six dimensions is given in section 4.  Finally, the
(Minkowski)$_3\times S^3$ vacuum is investigated in section 5, and we
end with a discussion on four-dimensional chirality in section 6.

\section{Supersymmetry of the generalized reduction}

The bosonic field content of half-maximum supergravity in seven
dimensions comprises a metric $\hat g_{MN}$, a scalar $\hat\phi$, an
antisymmetric tensor $\hat B_\2$ and three vectors $\hat A_\1^{a}$.
The Lagrangian in the bosonic sector is \cite{townsendvan,salsez2}
\begin{equation}
\hat {\cal L}=\hat R {\hat *\oneone} -\ft12{\hat *d\hat\phi}\wedge 
d\hat\phi 
-\ft1{2}e^{\fft4{\sqrt{10}}\hat\phi}{\hat *\hat H_\3}\wedge
\hat H_\3
-\ft12e^{\fft2{\sqrt{10}}\hat\phi}{\hat *\hat F_\2^{a}}\wedge
\hat F_\2^a,
\end{equation}
where $\hat F_\2 ^{a}=d\hat A_\1^{a}$ and $\hat H_\3=
d\hat B_\2 -\fft12\hat F_\2^{a}\wedge\hat A_\1^{a}$.

\subsection{The supersymmetry transformations}

The fermionic sector consists of a pair of symplectic-Majorana gravitinos
$\hat\psi_{M\,i}$ as well as a pair of dilatinos $\hat\lambda_i$, where
$i=1,2$ is an ${\rm Sp}(1)$ index.  The three vectors form a triplet under
${\rm Sp}(1)$, and may equivalently be written as $\hat A_{\1\,i}{}^j
=\hat A_\1^{a}(-\tau^a)_i{}^j$ where $\tau^a$ are the usual
Pauli matrices.  In this form, the supersymmetry transformations on the
fermions are given by
\bea
\delta\hat\psi_{\sst M\,i}&=&[\hat\nabla_{\sst M}
-\ft1{60}(\hat\gamma_{\sst M}{}^{\sst{NPQ}}
-\ft92\delta_{\sst M}^{\sst N}\hat\gamma^{\sst{PQ}})
\,e^{\fft12\hat a\hat\phi}\hat H_{\sst{NPQ}}]\hat\epsilon_i\nn\\
&&+\,\ft{i}{20\sqrt2}(\hat\gamma_{\sst M}{}^{\sst{NP}}
-8\delta_{\sst M}^{\sst N}\hat\gamma^{\sst P})
e^{\fft14\hat a\hat\phi}\hat F_{\sst{NP}\,i}{}^j\hat\epsilon_j,\nn\\
\delta\hat\lambda_i&=&[-\ft1{2\sqrt2}\hat\gamma^{\sst M}
\partial_{\sst M}\hat\phi+\ft1{12\sqrt5}e^{\fft12\hat a\hat\phi}
\hat H_{\sst{MNP}}\hat\gamma^{\sst{MNP}}]
\hat\epsilon_i-\ft{i}{4\sqrt{10}}\,e^{\fft14\hat a\hat\phi}
\hat F_{\sst{MN}i}{}^j\hat\gamma^{\sst{MN}}\hat\epsilon_j,
\label{eq:f7susy}
\eea
where $\hat a=4/\sqrt{10}$.

In addition, the transformations on the bosonic fields have the form
\bea
\delta\hat\phi&=&-\ft1{2\sqrt2}\bar{\hat\epsilon^i}\hat\lambda_i,\nn\\
\delta\hat g_{\sst{MN}}&=&\ft12\bar{\hat\epsilon^i}
\hat\gamma_{(\sst M}\hat\psi_{\sst N)\,i},\nn\\
\delta\hat A_{\sst M\,i}{}^j&=&\ft{i}{\sqrt2}e^{-\fft14\hat a\hat\phi}
(\bar{\hat\psi^j}_{\sst M}-\ft1{\sqrt5}\bar{\hat\lambda^j}
\hat\gamma_{\sst M})\hat\epsilon_i,\nn\\
\delta\hat B_{\sst{MN}}&=&-\ft12\hat A_{[\sst M\,i}{}^j
\delta\hat A_{\sst N]\,j}{}^i-\ft12e^{-\fft12\hat a\hat\phi}
(\bar{\hat\psi^i}_{[\sst M}\hat\gamma_{\sst N]}
-\ft1{\sqrt5}\bar{\hat\lambda^i}\hat\gamma_{\sst{MN}})\hat\epsilon_i,
\label{eq:bsusy}
\eea
where in the transformation for $\hat A_{\sst M\,i}{}^j$,
the Sp(1) indices $i$ and $j$ are to be taken in the triplet
combination.  In particular, this may be enforced by the projection
$(\delta_i^{i'}\delta^j_{j'}-\fft12\delta_i^j \delta^{i'}_{j'})$
which removes the trace.  Note that the transformation for $\delta\hat
B_{\sst{MN}}$ is given in a dualized form compared to that of
\cite{townsendvan}.

The above fermionic (\ref{eq:f7susy}) and bosonic (\ref{eq:bsusy})
supersymmetries are normalized according to
\begin{equation}
[\delta_1,\delta_2]\hat\Xi=\ft14\hat\xi^{\sst M}\partial_{\sst M}\hat\Xi
+\hbox{(local Lorentz)}+\hbox{(general coordinate)}+\hbox{(gauge)},
\end{equation}
where $\hat\xi^{\sst M}=\bar{\hat\epsilon}{}_2^i\hat\gamma^{\sst M}\hat\epsilon_{1i}$.
Furthermore, when working with the fermions, it is often convenient to
make use of the Majorana flip conditions
\begin{eqnarray}
&&\bar{\hat\chi}{}^i\gamma_{\sst{M_1M_2\cdots M_n}}^{\vphantom{i}}\hat\psi_i=
(-)^n\bar{\hat\psi}{}^i\gamma_{\sst{M_nM_{n-1}
\cdots M_1}}^{\vphantom{i}}\hat\chi_i,\nn\\
&&\bar{\hat\chi}{}^j\gamma_{\sst{M_1M_2\cdots M_n}}^{\vphantom{i}}\hat\psi_i=
(-)^{n+1}\bar{\hat\psi}{}^j\gamma_{\sst{M_nM_{n-1}
\cdots M_1}}^{\vphantom{i}}\hat\chi_i,
\label{eq:mflip}
\end{eqnarray}
for the singlet and triplet combinations, respectively.

\subsection{The bosonic reduction ansatz}

As demonstrated in \cite{kerimo}, the generalized $S^1$ reduction ansatz is
given on the bosonic fields by
\begin{eqnarray}
d\hat s_7^2&=&e^{2m_2z}\left(e^{2\alpha\varphi}ds_6^2+e^{2\beta\varphi}
(dz+{\cal A}_\1)^2\right),\nn\\
\hat B_\2&=&e^{2(m_2-m_1)z}(B_\2+B_\1\wedge dz),\nn\\
\hat A_\1^{a}&=&e^{(m_2-m_1)z}(A_\1^{a}+\Phi^{a}dz),\nn\\
\hat\phi&=&\phi+\sqrt{10}\,m_1z,
\label{eq:bosred}
\end{eqnarray}
where $\alpha^2=\fft1{40}$ and $\beta=-4\alpha$.  The resulting reduction
yields the six-dimensional fields
$(g_{\mu\nu},\cA_\1,A_\1^{a},B_\2,\phi_1)$
and $(B_\1,\phi_2,\Phi^{a})$ corresponding to the bosonic content of
${\cal N}=(1,1)$ supergravity coupled to a vector multiplet.  Note that
the ${\rm Sp}(1)$ singlet graviphoton and the matter vector are in actuality
given by linear combinations of $\cA_\1$ and $B_\1$.  However, the
scalars $\phi_1$ and $\phi_2$, given by the rotated combinations
\bea
\phi_1=\ft2{\sqrt{5}}\phi-\ft1{\sqrt{5}}\varphi,\nn\\
\phi_2=\ft1{\sqrt{5}}\phi+\ft2{\sqrt{5}}\varphi,
\eea
are diagonal between multiplets.

\subsection{The fermionic reduction}

Working out the fermion reduction is straightforward, although somewhat
tedious.  Since the resulting $D=6$ theory contains a vector
multiplet in addition to the pure supergravity multiplet, the $D=7$
fermions $\hat\psi_{M\,i}$ and $\hat\lambda_i$ must reduce to yield
a $D=6$ gravitino and dilatino $(\psi_{\mu\,i},\lambda_i)$ as well as
a gaugino $\chi_i$.  The reduction from seven to six dimensions is
facilitated by the fact that the $D=7$ symplectic-Majorana condition
$\bar{\hat\psi}{}^i=-\epsilon^{ij}\hat\psi_j^T \hat C$ continues to apply 
in $D=6$, yielding a trivial reduction on the spinors.

Examination of the supersymmetry transformations on the fermions,
(\ref{eq:f7susy}), indicates that the proper fermionic reduction is given
by
\begin{eqnarray}
\hat\epsilon_i&=&e^{\fft12m_2z}e^{\fft12\alpha\varphi}\epsilon_i,\nonumber\\
\hat\lambda_i&=&\ft1{\sqrt{5}}e^{-\fft12m_2z}e^{-\fft12\alpha\varphi}
(\chi_i+2\lambda_i),\nonumber\\
\hat\psi_{z\,i}&=&\ft25e^{\fft12m_2z}e^{(\beta-\fft12\alpha)\varphi}\gamma_7
(2\chi_i-\lambda_i),\nonumber\\
\hat\psi_{\mu\,i}&=&e^{\fft12m_2z}e^{\fft12\alpha\varphi}[\psi_{\mu\,i}
+(\ft25e^{(\beta-\alpha)\varphi}\cA_\mu\gamma_7-\ft1{10}\gamma_\mu)
(2\chi_i-\lambda_i)].
\label{eq:fermred}
\end{eqnarray}
In this case, the resulting $D=6$ fermions have supersymmetry
transformations
\bea
\delta\psi_{\mu\,i}&=&[\nabla_\mu-\ft58m_2\gamma_\mu\gamma^\nu\cA_\nu
-\ft1{48}e^{\fft1{\sqrt{2}}\phi_1}(\gamma_\mu{}^{\nu\rho\sigma}
-3\delta_\mu^\nu\gamma^{\rho\sigma})H_{\nu\rho\sigma}
+\,\ft58m_2e^{\fft1{\sqrt{2}}\phi_2-\fft1{2\sqrt{2}}\phi_1}\gamma_\mu\gamma_7
\nn\\
&&\quad-\,\ft1{32}e^{\fft1{2\sqrt{2}}\phi_1}(\gamma_\mu{}^{\nu\rho}
-6\delta_\mu^\nu\gamma^\rho)\gamma_7(e^{\fft1{\sqrt{2}}\phi_2}H_{\nu\rho}
+e^{-\fft1{\sqrt{2}}\phi_2}\cF_{\nu\rho})]\epsilon_i\nn\\
&&\quad+[-\ft{i}{2\sqrt{2}}e^{\fft1{\sqrt{2}}\phi_2}\gamma_7Q_{\mu\,i}{}^j
+\ft{i}{16\sqrt{2}}e^{\fft1{2\sqrt{2}}\phi_1}(\gamma_\mu{}^{\nu\rho}
-6\delta_\mu^\nu\gamma^\rho)F_{\nu\rho\,i}{}^j]\epsilon_j,\nn\\
\delta\lambda_i&=&[-\ft1{2\sqrt{2}}\gamma^\mu\partial_\mu\phi_1+\ft14(4m_1+m_2)
\cA_\mu\gamma^\mu+\ft1{24}e^{\fft1{\sqrt{2}}\phi_1}H_{\mu\nu\rho}
\gamma^{\mu\nu\rho} \nn\\
&&\quad+\,\ft1{16}e^{\fft1{2\sqrt{2}}\phi_1}\gamma^{\mu\nu}
\gamma_7(e^{\fft1{\sqrt{2}}\phi_2}H_{\mu\nu}+e^{-\fft1{\sqrt{2}}
\phi_2}\cF_{\mu\nu})
-\ft14(4m_1+m_2)e^{\fft1{\sqrt{2}}\phi_2-\fft1{2\sqrt{2}}\phi_1}\gamma_7]
\epsilon_i\nn\\
&&\quad-\ft{i}{8\sqrt{2}}e^{\fft1{2\sqrt{2}}\phi_1}\gamma^{\mu\nu}
F_{\mu\nu\,i}{}^j\epsilon_j,\nn\\
\delta\chi_i&=&[-\ft1{2\sqrt{2}}\gamma^\mu\partial_\mu\phi_2
+\ft12(m_1-m_2)\cA_\mu\gamma^\mu
-\ft12(m_1-m_2)e^{\fft1{\sqrt{2}}\phi_2-\fft1{2\sqrt{2}}\phi_1}\gamma_7\nn\\
&&\quad-\ft18e^{\fft1{2\sqrt{2}}\phi_1}\gamma^{\mu\nu}
\gamma_7(-e^{\fft1{\sqrt{2}}\phi_2}H_{\mu\nu}+e^{-\fft1{\sqrt{2}}
\phi_2}\cF_{\mu\nu})]\epsilon_i
-\ft{i}{2\sqrt{2}}e^{\fft1{\sqrt{2}}\phi_2}\gamma^\mu\gamma_7Q_{\mu\,i}{}^j
\epsilon_j.
\label{eq:fsusy}
\eea
Here the $D=6$ field strengths are given by
\begin{eqnarray}
H_\3\!&=&\!dB_\2-\ft12F_\2^{a}\wedge A_\1^{a}-dB_\1\wedge
\cA_\1-2(m_2-m_1)B_\2\wedge\cA_\1+\ft12\Phi^{a}F_\2^{a}
\wedge\cA_\1,\nn\\
H_\2\!&=&\!dB_\1-\ft12\Phi^{a}F_\2^{a}+\ft12Q_\1^{a}\wedge
A_\1^{a}-\ft12\Phi^{a}Q_\1^{a}\wedge\cA_\1+2(m_2-m_1)B_\2,
\nn\\
F_\2^{a}\!&=&\!dA_\1^{a}-d\Phi^{a}\wedge\cA_\1+(m_2-m_1)
A_\1^{a}\wedge\cA_\1,\nn\\
Q_\1^{a}\!&=&\!d\Phi^{a}-(m_2-m_1)A_\1^{a},
\label{eq:genfs}
\end{eqnarray}
with $Q_{\1\,i}{}^j=Q_\1^{a}(-\tau^a)_{ij}$, {\it etc.}

The gravitino transformation in (\ref{eq:fsusy}) demonstrates that
the ${\rm Sp}(1)$ singlet graviphoton arises as a linear combination
of $H_{\mu\nu}$ and $\cF_{\mu\nu}$.  Note, further, that these
transformations reduce to those of ordinary ungauged ${\cal N}=(1,1)$
supergravity coupled to a vector multiplet in the limit of vanishing $m_1$
and $m_2$.  

\subsection{Generalized supersymmetry in six dimensions}

Given the bosonic (\ref{eq:bosred}) and fermionic (\ref{eq:fermred})
reductions, it is now a matter of substituting these expressions
into (\ref{eq:bsusy}) to obtain the $D=6$ bosonic transformations.
We find
\begin{eqnarray}
\delta\phi_1&=&-\ft1{2\sqrt{2}}\bar\epsilon^i\lambda_i,\nn\\
\delta\phi_2&=&-\ft1{2\sqrt{2}}\bar\epsilon^i\chi_i,\nn\\
\delta g_{\mu\nu}&=&\ft12\bar\epsilon^i\gamma_{(\mu}\psi_{\nu)\,i},\nn\\
\delta\cA_\mu&=&\ft14e^{-\fft1{2\sqrt{2}}\phi_1+\fft1{\sqrt{2}}\phi_2}
[\bar\epsilon^i\gamma_7(\psi_{\mu\,i}+\ft12\gamma_\mu\lambda_i)
+\bar\epsilon^i\gamma_\mu\gamma_7\chi_i],\nn\\
\delta A_{\mu\,i}{}^j&=&-\Phi_i{}^j\delta\cA_\mu-\ft{i}{\sqrt{2}}
e^{-\fft1{2\sqrt{2}}\phi_1}\bar\epsilon^j(\psi_{\mu\,i}
+\ft12\gamma_\mu\lambda_i),\nn\\
\delta\Phi_i{}^j&=&-\ft{i}{\sqrt{2}}e^{-\fft1{\sqrt{2}}\phi_2}
\bar\chi^j\gamma_7\epsilon_i,\label{eq:susy9}\\
\delta B_\mu&=&\ft14\Phi_i{}^j(\delta A_{\mu\,j}{}^i+\Phi_j{}^i\delta\cA_\mu)
-\ft14A_{\mu\,i}{}^j\delta\Phi_j{}^i\nn\\
&&+\ft14e^{-\fft1{2\sqrt{2}}\phi_1 -\fft1{\sqrt{2}}\phi_2}
[\bar\epsilon^i\gamma_7(\psi_{\mu\,i}+\ft12\gamma_\mu\lambda_i)
-\bar\epsilon^i\gamma_\mu\gamma_7\chi_i],\nn\\
\delta B_{\mu\nu}&=&-\ft12A_{[\mu\,i}{}^j\Phi_j{}^i\delta\cA_{\nu]}
-2B_{[\mu}\delta\cA_{\nu]}-\ft12A_{[\mu\,i}{}^j\delta A_{\nu]\,j}{}^i
-\ft12e^{-\fft1{\sqrt{2}}\phi_1}\bar\epsilon^i(\gamma_{[\mu}\psi_{\nu]\,i}
+\ft12\gamma_{\mu\nu}\lambda_i).\nn
\end{eqnarray}
This result, combined with (\ref{eq:fsusy}) yield the complete (lowest
order) supersymmetry transformations of the variant ${\cal N}=(1,1)$
supergravity coupled to a vector multiplet.  Note that in obtaining
(\ref{eq:fsusy}) and (\ref{eq:susy9}), it was crucial that the ansatz
(\ref{eq:fermred}) allowed a {\it consistent} reduction from seven to
six dimensions, in which the dependence on the $z$ coordinate
cancelled in the seven-dimensional transformation rules.  This
guarantees that the resulting six-dimensional supersymmetry
transformations are symmetries of the six-dimensional variant
supergravity.

   As noted in \cite{kerimo}, the vector multiplet may be truncated away by
setting $m_1=m_2$ as well as
\begin{equation}
\phi_2=0,\qquad \Phi_i{}^j=0,\qquad B_\mu=\cA_\mu=\ft1{\sqrt{2}}A_\mu,
\qquad\chi_i=0.
\label{eq:trq}
\end{equation}
In this case, the $D=6$ field strengths of (\ref{eq:genfs})
simplify to
\begin{eqnarray}
&&H_\3=dB_\2-\ft12F_{\2}^a\wedge A_{\1}^a -\ft12F_\2
\wedge A_\1,\nn\\
&&F_\2=dA_\1,\qquad F_{\2}^a=dA_{\1}^a .
\label{eq:fstr}
\end{eqnarray}
The resulting six dimensional theory has field content $(g_{\mu\nu},
A_\mu,A_\mu^{a},B_{\mu\nu},\phi_1,\psi_{\mu\,i},\lambda_i)$ and
supersymmetry transformations
\newpage
\bea
\delta\psi_{\mu\,i}&=&[\nabla_\mu-\ft5{8\sqrt{2}}m\gamma_\mu\gamma^\nu A_\nu
-\ft1{48}e^{\fft1{\sqrt{2}}\phi_1}(\gamma_\mu{}^{\nu\rho\sigma}
-3\delta_\mu^\nu\gamma^{\rho\sigma})H_{\nu\rho\sigma}\nn\\
&&\qquad-\,\ft1{16\sqrt{2}}e^{\fft1{2\sqrt{2}}\phi_1}(\gamma_\mu{}^{\nu\rho}
-6\delta_\mu^\nu\gamma^\rho)\gamma_7F_{\nu\rho}
+\ft58me^{-\fft1{2\sqrt{2}}\phi_1}\gamma_\mu\gamma_7]\epsilon_i\nn\\
&&+\,\ft{i}{16\sqrt{2}}e^{\fft1{2\sqrt{2}}\phi_1}(\gamma_\mu{}^{\nu\rho}
-6\delta_\mu^\nu\gamma^\rho)F_{\nu\rho\,i}{}^j\epsilon_j,\nn\\
\delta\lambda_i&=&[-\ft1{2\sqrt{2}}\gamma^\mu\partial_\mu\phi_1
+\ft5{4\sqrt{2}}m
A_\mu\gamma^\mu+\ft1{24}e^{\fft1{\sqrt{2}}\phi_1}H_{\mu\nu\rho}
\gamma^{\mu\nu\rho}\nn\\
&&\quad+\,\ft1{8\sqrt{2}}e^{\fft1{2\sqrt{2}}\phi_1}\gamma^{\mu\nu}
\gamma_7F_{\mu\nu}
-\ft54m e^{-\fft1{2\sqrt{2}}\phi_1}\gamma_7]\epsilon_i
-\,\ft{i}{8\sqrt{2}}e^{\fft1{2\sqrt{2}}\phi_1}\gamma^{\mu\nu}
F_{\mu\nu\,i}{}^j\epsilon_j,\nn\\
\delta\phi_1&=&-\ft1{2\sqrt{2}}\bar\epsilon^i\lambda_i,\nn\\
\delta g_{\mu\nu}&=&\ft12\bar\epsilon^i\gamma_{(\mu}\psi_{\nu)\,i},\nn\\
\delta A_\mu&=&\ft1{2\sqrt{2}}e^{-\fft1{2\sqrt{2}}\phi_1}
\bar\epsilon^i\gamma_7(\psi_{\mu\,i}+\ft12\gamma_\mu\lambda_i),\nn\\
\delta A_{\mu\,i}{}^j&=&-\ft{i}{\sqrt{2}}
e^{-\fft1{2\sqrt{2}}\phi_1}\bar\epsilon^j(\psi_{\mu\,i}+\ft12\gamma_\mu
\lambda_i),\nn\\
\delta B_{\mu\nu}&=&-A_{[\mu}\delta A_{\nu]}
-\ft12A_{[\mu\,i}{}^j\delta A_{\nu]\,j}{}^i
-\ft12e^{-\fft1{\sqrt{2}}\phi_1}\bar\epsilon^i(\gamma_{[\mu}\psi_{\nu]\,i}
+\ft12\gamma_{\mu\nu}\lambda_i).
\label{eq:tfsusy}
\eea
These transformations reduce to those of \cite{romans} when $m\to0$.

On the other hand, for $m\ne0$, the generalized reduction yields
additional terms in $\delta\psi_{\mu\,i}$ and $\delta\lambda_i$.
Furthermore, these $m$-dependent terms do not have the usual structure
for a gauged supergravity.  In particular, the gauge potential $A_\1$
does not appear in $\delta\psi_{\mu\,i}$ as a minimal coupling term
$D_\mu=\nabla_\mu +igA_\mu$ to a charged spinor, yet shows up as a
bare potential term in $\delta\lambda_i$.  This is consistent with
$A_\1$ showing up as well in the bosonic equations of motion
\cite{kerimo}.  For this reason, it is natural to suspect that the
local supersymmetry algebra satisfied by this theory is necessarily
modified.  To see this, we may examine, {\it e.g.}, the double
variation on $\phi_1$.  We find
\begin{equation}
[\delta_1,\delta_2]\phi_1=\ft14\xi^\mu\partial_\mu\phi_1-\ft{5}{4\sqrt{2}}
m \Bigl(\ft1{\sqrt{2}}\xi^\mu A_\mu
-e^{-\fft1{2\sqrt{2}}\phi_1}(\bar\epsilon_2^i\gamma_7
\epsilon_{1i})\Bigr),
\end{equation}
where $\xi^\mu=\bar\epsilon_2^i\gamma^\mu\epsilon_{1i}$.
The additional terms vanish when $m=0$.

\subsection{String frame formalism}

Some of the aspects of this theory becomes simpler to discuss
in the string frame.  The relation between the quantities in the
Einstein frame and string frame is given by
\bea
&&g_{\mu\nu} = e^{-\phi}\, \td g_{\mu\nu}\,,\qquad
F_\2^a=\wtd F_\2^a\,,\qquad B_\2=\wtd B_\2\,,\qquad
d\phi + \ft{5}{2\sqrt2}m A_\1 =\wtd B_\1\,,\nn\\
&&F_\2=\wtd F_\2\,,\quad \phi_1=-\sqrt2\,\phi\,,\quad
\epsilon=e^{-\fft14\phi}\, \td\epsilon\,,\quad
\lambda=e^{\fft14\phi}\, \td \lambda\,,\quad
\psi_\mu=e^{-\fft14\phi}\, \td\psi_\mu\,,
\eea
where the tilded variables are those in the string frame.  The bosonic
equations of motion are then given by
\bea
\wtd\nabla^{\sigma}\wtd H_{\mu\nu\sigma}&=&
2\wtd H_{\mu\nu\sigma}\wtd B^{\sigma}
-\ft{5}
{\sqrt2}\, m\, \wtd F_{\mu\nu}\,,\nn\\
\wtd \nabla^{\nu}\wtd F_{\mu\nu}&=&2\wtd F_{\mu\nu}\wtd B^{\nu}+\ft12
\wtd H_{\mu\nu\sigma} \wtd F^{\nu\sigma}\,,\nn\\
\wtd \nabla^{\nu}\wtd F^a_{\mu\nu}&=&2\wtd F^a_{\mu\nu}\wtd B^{\nu}+
\ft12\wtd H_{\mu\nu\sigma} \wtd F^{a\,\nu\sigma}\,,\nn\\
\wtd \nabla^{\mu}\wtd B_{\mu}&=&2\wtd B_\1^2-\ft1{12}\wtd H^{\,2}_\3-
\ft18(\wtd F^{\,2}_\2
+(\wtd F^{a}_\2)^2 )+\ft{25}2m^2\,,\nn\\
\wtd R_{\mu\nu}&=&-\wtd \nabla_{\mu}\wtd B_{\nu}-
\wtd \nabla_{\nu}\wtd B_{\mu}+\ft14\wtd H_{\mu\sigma\rho}
\wtd H_{\nu}^{\;\;\sigma\rho}+\ft12(\wtd F_{\mu\rho}\,\wtd F^{\;\,\rho}_{\nu}
+\wtd F^a_{\mu\rho}\,\wtd F^{a\;\rho}_{\nu})\,,
\eea
Thus we see that the dilaton $\phi_1$ is eaten by the gauge
field $A_\1$ to give rise to a massive vector $\wtd B_\1$ field.
The supersymmetric transformation rules for the fermions are given by
\bea
\delta\tilde\lambda_i&=&[\ft12\wtd B_{\mu}\,\tilde\gamma^{\mu}+\ft1{24}
\wtd H_{\mu\nu\rho}\tilde\gamma^{\mu\nu\rho}+
\ft1{8\sqrt2}\tilde\gamma^{\mu\nu}
\gamma_7\wtd F_{\mu\nu}-\ft54m\gamma_7]\tilde\epsilon_i
-\ft{i}{8\sqrt{2}}\tilde\gamma^{\mu\nu}\wtd F_{\mu\nu\,i}{}^j
\tilde\epsilon_j,
\nn\\
\delta\tilde\psi_{\mu\,i}&=&[\widetilde\nabla_\mu
-\ft14\wtd B_{\nu}\,\tilde\gamma_{\mu}\tilde\gamma^{\nu}
-\ft1{48}(\tilde\gamma_\mu{}^{\nu\rho\sigma}
-3\delta_\mu^\nu\tilde\gamma^{\rho\sigma})\wtd H_{\nu\rho\sigma}
+\ft58m\tilde\gamma_\mu\gamma_7\nn\\
&&-\,\ft1{16\sqrt2}(\tilde\gamma_\mu{}^{\nu\rho}-6\delta_\mu^\nu
\tilde\gamma^\rho)\gamma_7\wtd F_{\nu\rho}]\tilde\epsilon_i
+\,\ft{i}{16\sqrt2}(\tilde\gamma_\mu{}^{\nu\rho}-6\delta_\mu^\nu
\tilde\gamma^\rho)\wtd F_{\nu\rho\,i}{}^j\tilde\epsilon_j\,.
\eea
It is of interest to note that the supersymmetric transformation
rule for the shifted gravitino, $\wtd{\wtd\psi}_\mu
=\wtd \psi_\mu + \ft12 \td \gamma_\mu \td \lambda$, given by
\bea
\delta \wtd{\wtd\psi}_{\mu\,i}
=\Bigl[\wtd \nabla_\mu+\ft18\wtd H_{\mu\nu\rho}\tilde\gamma^{\nu\rho}
+\ft1{2\sqrt{2}}\tilde\gamma^\nu\gamma_7 \wtd F_{\mu\nu}\Bigr]\tilde\epsilon_i
-\ft{i}{2\sqrt{2}}\tilde\gamma^\nu \wtd F_{\mu\nu\,i}{}^j
\tilde\epsilon_j,
\eea
does not depend on $m$.

\section{The (Minkowski)$_4\times S^2$ reduction}

The $D=6$ theory obtained in \cite{kerimo} does not admit a Lagrangian
formulation since the bare potential $A_\1$ appears directly in the
equations of motion.  This is also apparent from the supersymmetry
variations obtained in the previous section.  However, for field
configurations with vanishing $A_\1$, the resulting bosonic equations
of motion may be obtained from the Lagrangian
\begin{equation}
{\cal L}=\hat R{\hat*\oneone}-\ft14{\hat*d\hat\phi}\wedge d\hat\phi
-\ft12e^{\hat\phi}{\hat*\hat H_\3}\wedge
\hat H_\3-\ft12e^{\fft12\hat\phi}{\hat*\hat F_\2^{a}}\wedge\hat F_\2^{a}
-8g^2e^{-\fft12\hat\phi}{\hat*\oneone}.
\label{eq:d6lag}
\end{equation}
We have now introduced carets to denote six-dimensional fields, in
anticipation of a subsequent reduction to four dimensions.
Furthermore, we have defined $\hat\phi=\sqrt{2}\, \phi_1$ to simplify
the subsequent expressions and have defined $5m=2\sqrt{2}\, g$.

Curiously, this bosonic Lagrangian is identical to that of the
Salam-Sezgin model, with the exception that there are three vector
fields instead of one.  As a result, this model clearly admits a
bosonic $M_4\times S^2$ reduction, where $M_4$ denotes
four-dimensional
Minkowski spacetime.  On the other hand, the
supersymmetry of the reduction must still be verified, as the
supersymmetry transformations of the variant ${\cal N}=(1,1)$ theory
differ from that of the gauged ${\cal N}=(1,0)$ model.

In order to investigate the supersymmetry, it is useful to rewrite the
six-dimensional symplectic-Majorana spinors using a Dirac notation.
A symplectic-Majorana spinor satisfies the reality condition 
$(\hat\psi_i)^*=-\epsilon^{ij}\hat C\hat\gamma_0\hat\psi_j$, where
the charge conjugation matrix $\hat C$ satisfies $\hat C^T = \hat C$
and $\hat C^\dagger\, \hat C=1$.  We may now 
form the Dirac combination $\hat\psi=\hat\psi_1+i\hat\psi_2$, with
complex conjugate
$\hat\psi^*=-i\hat C\hat\gamma_0(\hat\psi_1-i\hat\psi_2)$.  Equivalently,
these definitions may be inverted to yield
\begin{equation}
\hat\psi_1=\ft12(\hat\psi-i\hat\gamma_0\hat C^*\hat\psi^*),\qquad
\hat\psi_2=\ft1{2i}(\hat\psi+i\hat\gamma_0\hat C^*\hat\psi^*).
\end{equation}
As a result, for $\hat A_\1=0$, the supersymmetry transformations
(\ref{eq:tfsusy}) may be rewritten as
\bea
\delta\hat\psi_\mu&=&\Bigl[\hat\nabla_\mu
-\ft1{48}e^{\fft12\hat\phi}(\hat\gamma_\mu{}^{\nu\rho\sigma}
-3\delta_\mu^\nu\hat\gamma^{\rho\sigma})\hat H_{\nu\rho\sigma}
+\ft1{2\sqrt{2}}ge^{-\fft14\hat\phi}\hat\gamma_\mu\hat\gamma_7
\Bigr]
\hat\epsilon\nn\\
&&\qquad+\,\ft{i}{16\sqrt{2}}e^{\fft14\hat\phi}(\hat\gamma_\mu{}^{\nu\rho}
-6\delta_\mu^\nu\hat\gamma^\rho)\Bigl(\hat F_{\nu\rho}^{2}\hat\epsilon
-(\hat F_{\nu\rho}^{1}-i\hat F_{\nu\rho}^{3})\hat\gamma_0\hat C^*
\hat\epsilon^*\Bigr),\nn\\
\delta\hat\lambda&=&\Bigl[-\ft14\hat\gamma^\mu\partial_\mu\hat\phi
+\ft1{24}e^{\fft12\hat\phi}\hat H_{\mu\nu\rho}\hat\gamma^{\mu\nu\rho}
-\ft1{\sqrt{2}}g e^{-\fft14\hat\phi}\hat\gamma_7\Bigr]\hat\epsilon \nn\\
&&\qquad-\,\ft{i}{8\sqrt{2}}e^{\fft14\hat\phi}\hat\gamma^{\mu\nu}
\Bigl(\hat F_{\mu\nu}^{2}\hat\epsilon-(\hat F_{\mu\nu}^{1}-i\hat
F_{\mu\nu}^{3})\hat\gamma_0\hat C^*\hat\epsilon^*\Bigr),
\label{eq:d6fsusy}
\eea
for the fermions, and
\bea
\delta\hat\phi&=&-\ft14[\bar{\hat\epsilon}\hat\lambda+\bar{\hat\lambda}
\hat\epsilon],\nn\\
\delta \hat g_{\mu\nu}&=&\ft12[\bar{\hat\epsilon}\hat\gamma_{(\mu}
\hat\psi_{\nu)}-\bar{\hat\psi}_{(\mu}\hat\gamma_{\nu)}\hat\epsilon],\nn\\
\delta \hat A_\mu&=&\ft1{4\sqrt{2}}e^{-\fft14\hat\phi}
[\bar{\hat\epsilon}\hat\gamma_7(\hat\psi_{\mu}+\ft12\hat\gamma_\mu
\hat\lambda)-(\bar{\hat\psi_\mu}-\ft12\bar{\hat\lambda}\hat\gamma_\mu)
\hat\gamma_7\hat\epsilon],\nn\\
\delta\hat A_\mu^1&=&-\ft1{2\sqrt{2}}e^{-\fft14\hat\phi}
\Im[\hat\epsilon^T\hat C(\hat\psi_\mu+\ft12\hat\gamma_\mu\hat\lambda)],\nn\\
\delta\hat A_\mu^2&=&-\ft{i}{4\sqrt{2}}e^{-\fft14\hat\phi}
[\bar{\hat\epsilon}(\hat\psi_\mu+\ft12\hat\gamma_\mu\hat\lambda)
-(\bar{\hat\psi}_\mu-\ft12\bar{\hat\lambda}\hat\gamma_\mu)\hat\epsilon],\nn\\
\delta\hat A_\mu^3&=&-\ft1{2\sqrt{2}}e^{-\fft14\hat\phi}
\Re[\hat\epsilon^T\hat C(\hat\psi_\mu+\ft12\hat\gamma_\mu\hat\lambda)],\nn\\
\delta \hat B_{\mu\nu}&=&
-\hat A_{[\mu}^{a}\delta\hat A_{\nu]}^{a}
-\ft14e^{-\fft12\hat\phi}[\bar{\hat\epsilon}(\hat\gamma_{[\mu}\hat\psi_{\nu]}
+\ft12\hat\gamma_{\mu\nu}\hat\lambda)
+(\bar{\hat\psi}_{[\mu}\hat\gamma_{\nu]}-\ft12\bar{\hat\lambda}
\hat\gamma_{\mu\nu})\hat\epsilon],
\label{eq:d6bsusy}
\eea
for the bosons.  While we have set $\hat A_\mu =0$, it is important to
retain its supersymmetry variation so that it is possible to check
later for consistency.  These expressions serve as the starting point
for the subsequent analysis.

\subsection{Supersymmetry of the $M_4\times S^2$ vacuum}

The bosonic theory, given by (\ref{eq:d6lag}), admits an $M_4\times S^2$
solution given by
\begin{eqnarray}
&&d\hat s_6^2=\eta_{\mu\nu}dx^\mu dx^\nu+\fft1{8g^2}d\Omega_2^2,\nn\\
&&\hat F_\2^2=\fft1{2g}\Omega_{(2)},
\label{eq:m4s2vac}
\end{eqnarray}
where $\Omega_{(2)}=\sin\theta\,d\theta\wedge d\varphi$ is the volume form on the
unit $S^2$.  Note that we have singled out the $2$-component of the
${\rm Sp}(1)$ triplet gauge fields for convenience.  While this choice is
a natural one corresponding to the Dirac combination in (\ref{eq:d6fsusy}),
any other choice would yield the same result.

To examine the supersymmetry of the vacuum, we insert (\ref{eq:m4s2vac})
into (\ref{eq:d6fsusy}) to obtain
\begin{eqnarray}
\delta\hat\psi_\alpha&=&[\partial_\alpha+\ft1{\sqrt{2}}g\hat\gamma_\alpha
\hat\gamma_7P_+]\hat\epsilon,\nn\\
\delta\hat\psi_a&=&[\nabla_a-i\sqrt{2}g\hat\gamma_a\hat\gamma_{45}
]\hat\epsilon+\ft1{\sqrt{2}}g\hat\gamma_a\hat\gamma_7P_+\hat\epsilon,\nn\\
\delta\hat\lambda&=&-\sqrt{2}g \hat\gamma_7P_+\hat\epsilon
\end{eqnarray}
where $P_\pm=\fft12(1\pm i\hat\gamma^{45} \hat\gamma_7)$ is a half-BPS
projection.  These equations vanish for $\hat\epsilon=P_-\hat\epsilon_0$
where $\hat\epsilon_0$ solves the Killing spinor equation on the round
2-sphere,
$[\nabla_a-i\sqrt{2}g\hat\gamma_a\hat\gamma_{45}]\hat\epsilon_0=0$.

To be more precise, we decompose the six-dimensional Dirac matrices
according to
\begin{eqnarray}
&&\hat\gamma_\alpha=\gamma_\alpha\otimes\sigma_3,\qquad
\hat\gamma_4=\oneone\otimes\sigma_1,\qquad
\hat\gamma_5=\oneone\otimes\sigma_2,\nn\\
&&\hat\gamma_7=\hat\gamma_0\hat\gamma_1\cdots\hat\gamma_5=\gamma^5\otimes
\sigma_3,\qquad\hat C=C\otimes\sigma_2
\label{eq:dirdec}
\end{eqnarray}
where $C$ is now the four-dimensional charge conjugation matrix and
$\gamma^5=i\gamma_0 \gamma_1\gamma_2\gamma_3$.  Six dimensional
spinors $\hat\epsilon$ may then be written in terms of $M_4$ and $S^2$
spinors as $\hat\epsilon=\sum_I\epsilon_I\otimes\eta_I$ where $\eta_I$
is taken to be commuting.  In this case, the Killing spinor equation on
$S^2$ becomes $[\nabla_a+\sqrt{2}g\sigma_a\sigma_3] \eta_I=0$, and yields
two independent solutions.  Corresponding to the above choice of Dirac
matrices, we find that in the basis $e^4 = (2\sqrt 2 g)^{-1}\, d\theta$,
$e^5 = (2\sqrt2 g)^{-1}\, \sin\theta\, d\varphi$, the two independent
Killing spinors can be written as
\be
\eta_1 = \pmatrix{\cos\ft12\theta\cr -\sin\ft12\theta}\, e^{\ft{\im}{2}\,
\varphi}\,,\qquad
\eta_2 = \pmatrix{\sin\ft12\theta\cr \cos\ft12\theta}\, e^{-\ft{\im}{2}\,
\varphi}\,.
\ee
It is easily seen that these satisfy the conditions
\begin{equation}
\bar\eta_I\eta_J=\delta_{IJ},\qquad\eta_I^T\sigma^2\eta_J=-i\epsilon_{IJ},
\qquad \eta_I^*=i\sigma^2\epsilon_{IJ}\eta_J.
\label{eq:etacond}
\end{equation}
Note that $\bar\eta_I\equiv\eta_I^\dagger$.
Using the decomposition (\ref{eq:dirdec}), the half-BPS projection operator
takes the form $P_\pm=\fft12(1\mp\gamma_5)$.  As a result, the Killing
spinors in the $M_4\times S^2$ background are given by
\begin{equation}
\hat\epsilon=\epsilon_I\otimes\eta_I\qquad (\epsilon_I=\gamma_5\epsilon_I),
\end{equation}
where the $\epsilon_I$ are a pair of constant $D=4$ Weyl spinors.

\subsection{Reduction to $D=4$, ${\cal N}=2$ supergravity}

The existence of a supersymmetric vacuum suggests that a consistent
Kaluza-Klein reduction on $S^2$ is possible, yielding a Poincar\'e theory
in four dimensions.  Since the six-dimensional ${\cal N}=(1,1)$ theory has
16 real supersymmetries, and the vacuum breaks exactly half of them, the
resulting theory corresponds to ${\cal N}=2$ supersymmetry in four
dimensions.

The basic ${\cal N}=2$ supergravity multiplet consists of
a graviton $g_{\mu\nu}$, graviphoton $A_\1$ and a pair of Majorana
gravitinos $\psi_{\mu\,i}$.  In addition, ${\cal N}=2$ vector multiplets
are given by a vector $A_\1$, two real scalars $\phi$ and $a$, and a pair
of Majorana gauginos $\chi_i$.  We find that the six dimensional field
content reduces to yield ${\cal N}=2$ supergravity coupled to a single
vector multiplet.  The reduction ansatz for the bosons is given by
\begin{eqnarray}
&&d\hat s_6^2=e^{\fft12\phi}ds_4^2+\fft1{8g^2}e^{-\fft12\phi}d\Omega_2^2,\nn\\
&&\hat F_\2^2=2ge^{\fft12\phi}\epsilon_{ab}\hat e^a\wedge\hat e^b,\qquad
\hat F_\2^1=F_\2^1,\qquad \hat F_\2^3=F_\2^3,\nn\\
&&\hat H_\3=H_\3,\qquad\hat\phi=-\phi.
\label{eq:bosred4}
\end{eqnarray}
Note that the graviphoton and matter vector field strengths are given by
a combination of $F_\2^1$ and $F_\2^3$ (up to duality) as will be apparent
below.  The use of the $1$- and $3$-components of the ${\rm Sp}(1)$
triplet in the Kaluza-Klein reduction is dictated by the choice of turning
on $F_\2^2$ flux on the sphere.

It is straightforward to verify the consistency of the bosonic reduction.
The resulting
four-dimensional equations of motion may be obtained from the Lagrangian
\begin{equation}
{\cal L}=R{*\oneone}-\ft12{*d\phi}\wedge d\phi
-\ft12e^{-2\phi}{*H_\3}\wedge H_\3-\ft12e^{-\phi}
({*F_\2^1}\wedge F_\2^1+{*F_\2^3}\wedge F_\2^3).
\label{eq:d4lag}
\end{equation}

The fermion reduction ansatz may be obtained by substituting the
bosonic fields (\ref{eq:bosred4}) into the six-dimensional gravitino
and dilatino transformations (\ref{eq:d6fsusy}).  Starting with the
latter, we see that
\begin{eqnarray}
\delta\hat\lambda&=&
\sqrt{2}ge^{\fft14\phi}P_+\otimes\sigma_3\hat\epsilon
+e^{-\fft14\phi}[\ft14\gamma^\mu\partial_\mu\phi
+\ft1{24}e^{-\phi}H_{\mu\nu\rho}\gamma^{\mu\nu\rho}]\otimes\sigma_3\hat\epsilon
\nn\\
&&\qquad
-\ft{i}{8\sqrt{2}}e^{-\fft14\phi}[e^{-\fft12\phi}(F_{\mu\nu}^1-iF_{\mu\nu}^3)
\gamma^{\mu\nu}\gamma_0C^*]\otimes\sigma_3\sigma_2\hat\epsilon^*.
\label{eq:dilvar4}
\end{eqnarray}
The first term vanishes on chiral spinors $P_+\hat\epsilon=0$, while the
remaining terms combine to yield the four-dimensional gaugino transformation.

Turning to the gravitino variation, as usual the $D=6$ variation splits into
a $D=4$ gravitino term, $\delta\hat\psi_\alpha$, as well as two internal
variations, $\delta\hat\psi_a$.  Since the $S^2$ symmetry is unbroken by
the bosonic ansatz, the two internal components of the gravitino variation
are related by symmetry.  In fact, provided $\hat\epsilon$ is decomposed
in terms of Killing spinors on the sphere, the $\delta\hat\psi_a$ variation
has identical content as that of $\delta\hat\lambda$.  (This is not in
general true, but holds in the present case.)  As a result, we find the
fermionic reduction ansatz to have the form
\begin{eqnarray}
\hat\epsilon&=&e^{\fft18\phi}\epsilon_I\otimes\eta_I,\nn\\
\hat\lambda&=&e^{-\fft18\phi}\chi_I\otimes\sigma_3\eta_I,\nn\\
\hat\psi_\alpha&=&e^{-\fft18\phi}[\psi_{\alpha\,I}+\ft12\gamma_\alpha
\chi_I]\otimes\eta_I,\qquad
\hat\psi_a=e^{-\fft18\phi}(-\ft12\chi_I)\otimes\sigma_a\sigma_3\eta_I.
\end{eqnarray}

Inserting this ansatz into (\ref{eq:dilvar4}) as well as the gravitino
variations yields the four-dimensional supersymmetry transformations
\begin{eqnarray}
\delta\chi_I&=&[\ft14\gamma^\mu\partial_\mu\phi+\ft1{24}e^{-\phi}
H_{\mu\nu\rho}\gamma^{\mu\nu\rho}]\epsilon_I
-\ft1{4\sqrt{2}}e^{-\fft12\phi}(F_{\mu\nu}^1-iF_{\mu\nu}^3)\gamma^{\mu\nu}
\gamma_0C^*\epsilon_{IJ}\epsilon_J^*,\nn\\
\delta\psi_{\mu\,I}&=&[\nabla_\mu-\ft1{24}e^{-\phi}\gamma_\mu{}^{\nu\rho\sigma}
H_{\nu\rho\sigma}]\epsilon_I-\ft1{8\sqrt{2}}e^{-\fft12\phi}(F_{\nu\rho}^1
-iF_{\nu\rho}^3)\gamma^{\nu\rho}\gamma_\mu\gamma_0C^*\epsilon_{IJ}
\epsilon_J^*.\qquad
\end{eqnarray}
To obtain this result, we had to make use of the $\eta_I^*$ relation in
(\ref{eq:etacond}).  At this stage, we note that the gauge fields may
be dualized in four dimensions, so that $F_{\mu\nu}\gamma^{\mu\nu}
=-i{*F}_{\mu\nu}\gamma^{\mu\nu}\gamma_5$.  Since the four-dimensional
spinors are given in a Weyl basis
\begin{equation}
P_+\, \epsilon_I=0,\qquad P_+\, \psi_{\alpha\,I}=0,\qquad P_-\, \chi_I=0,
\label{eq:weylbasis}
\end{equation}
where $P_\pm=\fft12(1\mp\gamma_5)$, the above supersymmetry variations may
be rewritten as
\begin{eqnarray}
\delta\chi_I&=&[\ft14\gamma^\mu\partial_\mu\phi+\ft1{24}e^{-\phi}
H_{\mu\nu\rho}\gamma^{\mu\nu\rho}]\epsilon_I
-\ft1{4\sqrt{2}}e^{-\fft12\phi}(F_{\mu\nu}^1+*F_{\mu\nu}^3)\gamma^{\mu\nu}
\gamma_0C^*\epsilon_{IJ}\epsilon_J^*,\nn\\
\delta\psi_{\mu\,I}&=&[\nabla_\mu-\ft1{24}e^{-\phi}\gamma_\mu{}^{\nu\rho\sigma}
H_{\nu\rho\sigma}]\epsilon_I-\ft1{8\sqrt{2}}e^{-\fft12\phi}(F_{\nu\rho}^1
-*F_{\nu\rho}^3)\gamma^{\nu\rho}\gamma_\mu\gamma_0C^*\epsilon_{IJ}
\epsilon_J^*.\qquad
\end{eqnarray}
This highlights the nature of the ${\cal N}=2$ graviphoton,
$F_\2^{({\cal N}=2)}=e^{-\ft12\phi}F_\2^1+e^{\ft12\phi}\wtd F_\2^3$,
where $\wtd F_\2^3=e^{-\phi}{*F}_\2^3$.

Having completed the fermion reduction and supersymmetry variations, we now
turn to the reduction of the bosonic variations, (\ref{eq:d6bsusy}).  The
six-dimensional dilaton variation $\delta\hat\phi$ readily yields
$\delta\phi=\fft12\bar\epsilon_I\chi_I$.  Similarly, the four-dimensional
components of $\delta\hat g_{\mu\nu}$ yield $\delta g_{\mu\nu}=\fft12
\bar\epsilon_I\gamma_{(\mu}\psi_{\nu)\,I}$, while the internal components
reduce to give the identical $\delta\phi$ transformation.  This is a result
of setting the internal components of the six-dimensional gravitino equal
to the dilatino in the reduction.

In general, one obtains non-trivial vector field variations from the
mixed components of the metric, $\delta\hat g_{\mu i}$, as well as
directly from $\delta\hat A_\mu$.  However, these terms vanish identically
based on the $P_\pm$ chiralities of the four-dimensional spinors.  Likewise,
$\delta\hat A_\mu^2$ vanishes for the same reason.  On the other hand, the
additional complex conjugation appearing in $\delta\hat A_\mu^1$ and
$\delta\hat A_\mu^3$ prevents these transformations from vanishing.  The
resulting four-dimensional variations then have the form
\begin{eqnarray}
\delta g_{\mu\nu}&=&\ft14[\bar\epsilon_I\gamma_{(\mu}\psi_{\nu)\,I}
-\bar\psi_{(\mu\,I}\gamma_{\nu)}\epsilon_I],\nn\\
\delta \phi&=&\ft14[\bar\epsilon_I\chi_I+\bar\chi_I\epsilon_I],\nn\\
\delta B_{\mu\nu}&=&-\ft14e^\phi[\bar\epsilon_I\gamma_{[\mu}\psi_{\nu]\,I}
+\bar\psi_{[\mu\,I}\gamma_{\nu]}\epsilon_I+\bar\epsilon_I\gamma_{\mu\nu}
\chi_I-\bar\chi_I\gamma_{\mu\nu}\epsilon_I],\nn\\
\delta A_\mu^1&=&\ft1{2\sqrt{2}}e^{\fft12\phi}\epsilon_{IJ}\Re[\epsilon_I^TC
(\psi_{\mu\,J}+\ft12\gamma_\mu\chi_J)],\nn\\
\delta A_\mu^3&=&-\ft1{2\sqrt{2}}e^{\fft12\phi}\epsilon_{IJ}\Im[\epsilon_I^TC
(\psi_{\mu\,J}+\ft12\gamma_\mu\chi_J)].
\end{eqnarray}
We have verified that all variations of fields initially set to zero
vanish, either identically or through four-dimensional chirality.  This
verifies the consistency of the supersymmetric reduction to ${\cal N}=2$
supergravity coupled to a single vector multiplet.

\subsection{Truncation to $D=4$, ${\cal N}=1$ supergravity}

While we have retained ${\cal N}=2$ supersymmetry in the above
reduction, there is a natural truncation to ${\cal N}=1$.  This may
be accomplished by removing one of the two supersymmetry parameters by
setting $\epsilon_I =\hat n_I\epsilon$ where $\hat n_I$ is any constant
unit vector.  At the same time, it is necessary to truncate the ${\cal
N}=1$ gravitino and vector multiplets, leaving ${\cal N}=1$ supergravity
coupled to a chiral multiplet.  In the bosonic sector, this corresponds to
setting $A_\mu^1=A_\mu^3=0$.  The resulting bosonic Lagrangian is given by
\begin{equation}
{\cal L}=R{*\oneone}-\ft12{*d\phi}\wedge d\phi-\ft12e^{-2\phi}{*H_\3}
\wedge H_\3,
\end{equation}
while the relevant supersymmetry transformations are
\begin{eqnarray}
\delta\chi&=&[\ft14\gamma^\mu\partial_\mu\phi+\ft1{24}e^{-\phi}
H_{\mu\nu\rho}\gamma^{\mu\nu\rho}]\epsilon,\nn\\
\delta\psi_\mu&=&[\nabla_\mu-\ft1{24}e^{-\phi}\gamma_\mu{}^{\nu\rho\sigma}
H_{\nu\rho\sigma}]\epsilon,\nn\\
\delta g_{\mu\nu}&=&\ft14[\bar\epsilon\gamma_{(\mu}\psi_{\nu)}-\bar\psi_{(\mu}
\gamma_{\nu)}\epsilon],\nn\\
\delta\phi&=&\ft14[\bar\epsilon\chi+\bar\chi\epsilon],\nn\\
\delta B_{\mu\nu}&=&-\ft14e^{\phi}[\bar\epsilon\gamma_{[\mu}\psi_{\nu]}
+\bar\psi_{[\mu}\gamma_{\nu]}\epsilon+\bar\epsilon\gamma_{\mu\nu}\chi
-\bar\chi\gamma_{\mu\nu}\epsilon].
\end{eqnarray}

\section{BPS solutions}

The bosonic Lagrangian (\ref{eq:d4lag}) admits a dyonic black hole
solution where $F_\2^1$ is electric and $F_\2^3$ is magnetic (or vice
versa).  The solution is given by
\bea
ds_4^2 &=&- (\cH_1\, \cH_3)^{-1}\, dt^2 + \cH_1\, \cH_3 (dr^2 + r^2\,
d\wtd\Omega_2^2)
\,,\nn\\
F_\2^1 &=& dt\wedge d\cH_1^{-1}\,,\qquad  
F_\2^3 = q_3\, \wtd\Omega_\2 \,,\nn\\
\phi&=&-\log(\cH_1/\cH_3)\,,
\eea
where $\cH_1=1 + q_1/r$ and $\cH_3=1 + q_3/r$ are two harmonic
functions in the Euclidean three-dimensional transverse space.  It
becomes the standard Reissner-Nordstr\"om black hole when
$\cH_1=\cH_3$.  We can easily lift the solution back to $D=6$
dimensions, and it becomes
\bea
ds_6^2 &=& (\cH_1/\cH_3)^{\ft12}\,\Bigl[
-\cH_1^{-2}\, dt^2 + \cH_3^2\, (dr^2 + r^2\, d\wtd\Omega_2^2)
+ \fft1{8g^2}\,d\Omega_2^2\Bigr]\,,\nn\\
\hat F_\2^2 &=& \ft12 g^{-1}\, \Omega_\2\,,\qquad
\hat F_\2^1 = dt\wedge d\cH_1^{-1}\,,\qquad
\hat F_\2^3 = q_3\, \wtd\Omega_\2 \,,\nn\\
\hat \phi &=& \log(\cH_1/\cH_3)\,.
\eea
In the near horizon limit, the geometry becomes AdS$_2\times S^2\times
S^2$.  For $\cH_1=\cH_3$, the metric is the direct product of an $S^2$
and the Reissner-Nordstr\"om black hole.  In the string frame, the
metric is given by
\be
ds_{\rm str}^2 =
-\cH_1^{-2}\, dt^2 + \cH_3^2\, (dr^2 + r^2\, d\wtd\Omega_2^2)
+ \fft1{8 g^2}\, d\Omega_2^2
\ee

\section{(Minkowski)$_3\times S^3$ vacuum}

    The variant ${\cal N}=(1,1)$ six-dimensional supergravity has the unusual
feature that it admits not only a supersymmetric (Minkowski)$_4\times S^2$
vacuum, but also a supersymmetric (Minkowski)$_3\times S^3$ vacuum.  This
is quite different from the situation in the Salam-Sezgin theory; although
the Salam-Sezgin model admits a (Minkowski)$_3\times S^3$ solution as well as
a supersymmetric (Minkowski)$_4\times S^2$ solution, the former is 
non-supersymmetric.

   To construct the supersymmetric (Minkowski)$_3\times S^3$ solution in the
variant ${\cal N}=(1,1)$ supergravity, we make a standard Freund-Rubin type
ansatz in which
\be
d\hat s_6^2 = dx^\mu \, dx_\mu + ds_3^2\,,\qquad \hat H_\3 = q\, \ep_\3\,,
\qquad \hat\phi=0\,,
\ee
where $ds_3^2$ is the metric on a round $S^3$, with volume form $\ep_\3$,
and all other fields are set to zero.  We find that this solves the 
six-dimensional equations of motion if
\be
q = 2\sqrt2\, g\,.
\ee
The $S^3$ metric has Ricci tensor given by $R_{ij} = 4g^2\, g_{ij}$.  

    To establish the supersymmetry of the solution, we decompose the
six-dimensional Dirac matrices as
\be
\hat\gamma_\mu = \gamma_\mu\otimes\oneone \otimes \sigma_2\,,\quad
\hat\gamma_i = \oneone\otimes \gamma_i\otimes \sigma_1\,,\quad
\hat\gamma_7 = \oneone\otimes\oneone\otimes\sigma_3\,.
\ee
Writing $\hat\ep=\ep\otimes\eta\otimes\nu$, we find from the transformation 
rules (\ref{eq:d6fsusy}) that supersymmetry is preserved if $\ep$ is 
a constant spinor in the (Minkowski)$_3$ spacetime, $\sigma_2\, \nu=\nu$ and
if $\eta$ is a Killing spinor on $S^3$, satisfying
\be
\nabla_i\, \eta = \fft{\im\, g}{\sqrt2}\, \gamma_i\, \eta\,.
\ee
Thus the solution has three-dimensional ${\cal N}=4$ supersymmetry.

\section{Discussion}

It was shown in \cite{kerimo} that a variant ${\cal N}=(1,1)$ supergravity
may be constructed based on a generalized Scherk-Schwarz reduction from
seven dimensions.  In this paper, we have completed the analysis of
the fermion sector of this model, and have presented the variant
supersymmetry transformations in section 2.  In general, the resulting
theory contains two mass parameters, $m_1$ and $m_2$, and consists of a
single vector multiplet coupled to gravity.  A further truncation to pure
${\cal N}=(1,1)$ supergravity may be obtained by setting $m_1=m_2$.  

It may be seen from (\ref{eq:tfsusy}) that the resulting model reduces
to ordinary ${\cal N}=(1,1)$ ungauged supergravity \cite{romans} in the
limit $m\to0$.  In fact, the parameter $m$ is similar to a gauging
parameter in the sense that a potential $V=25m^2e^{-\fft12\hat\phi}$
is generated whenever $m\ne0$.  Nevertheless, this new theory is rather
unusual in that the bare vector potential terms in (\ref{eq:tfsusy})
do not correspond to the usual minimal coupling to charged fermions.  In
this sense, the ${\cal N}=(1,1)$ theory constructed in \cite{kerimo} differs
from conventional models with gauged $R$-symmetry.

In the bosonic sector (with vanishing ${\rm Sp(1)}$ singlet gauge field),
the field content and equations of motion of the variant ${\cal N}=(1,1)$
theory resembles that of the Salam-Sezgin model.  In particular, the
bosonic Lagrangian (\ref{eq:d6lag}) is identical to that of the gauged
${\cal N}=(1,0)$ model, albeit with a triplet of gauge fields.  Recall,
however, that the ${\cal N}=(1,1)$ supergravity multiplet decomposes into
a graviton, gravitino, and a tensor multiplet of ${\cal N}=(1,0)$
supersymmetry.  Thus the singlet and triplet gauge fields of the
${\cal N}=(1,1)$ supergravity reside in the gravitino multiplet, and not
a vector multiplet, as would be necessary for obtaining a Salam-Sezgin
truncation.  One practical implication of this observation is that, while
the present model admits similar vacuum solutions to the Salam-Sezgin
model, its supersymmetry properties could in principle be drastically
different.

As an example, we have demonstrated that the variant supergravity admits
an $M_4\times S^2$ reduction yielding $D=4$, ${\cal N}=2$ supergravity
coupled to a single vector multiplet.  Although we have used a Weyl notation
for the four-dimensional fermions, it is clear that this theory is non-chiral
in the usual sense.  Thus, while a further truncation to ${\cal N}=1$
supergravity coupled to a chiral multiplet is possible, we are unable to
obtain a chiral theory in four dimensions through this reduction process.

At this point, it is worth recalling that the presence of Weyl fermions in
four dimensions is insufficient to ensure a chiral theory.  Since
${\cal N}=1$ graviton and vector multiplets are inherently non-chiral,
it is necessary to have chiral multiplets transforming under a complex
representation of the gauge group in order to obtain a chiral model.
Note, in particular, that uncharged fermions may be described in either
Weyl or Majorana notation, so the presence of `left-handed' gravitinos
in (\ref{eq:weylbasis}) is not an indication of actual chirality.

While the present reduction does not result in massless charged fermions
in four dimensions, it should be emphasized that the $M_4\times S^2$
reduction of the Salam-Sezgin model likewise is non-chiral, although
the argument is somewhat subtle.  As demonstrated in \cite{witten},
smooth Kaluza-Klein reductions in the gravitational sector cannot lead
to a chiral theory in four dimensions.  However, \cite{witten} goes on to
indicate that chirality may be obtained by starting with chiral fermions
coupled to gauge fields in the higher dimensional theory, provided
the gauge reduction is non-trivial.  In particular, reductions with a
Monopole flux such as \cite{salamstrathdee} could in principle give rise
to four dimensional chirality.  This would suggest that the Salam-Sezgin
model is chiral, since it precisely involves turning on such a $U(1)$
monopole flux, with all fermions charged under this $U(1)$.  However, as
emphasized in \cite{quevedo1,gibbonpopes2}, the $U(1)$ does not survive
the reduction to four dimensions.  The resulting theory contains only
$SU(2)$ gauge fields and uncharged fermions, and is hence non-chiral.

Of course, the minimal Salam-Sezgin model in itself is anomalous, and
additional ${\cal N}=(1,0)$ vector and matter multiplets must be added to
cure the anomalies.  This presents a natural opportunity to construct a
model with six-dimensional charged chiral fermions.  Unfortunately, however,
to preserve supersymmetry the monopole flux used in the $M_4\times S^2$
reduction must be that corresponding to the gauging of a $U(1)$ subgroup of
${\rm Sp(1)}$.  Since it is precisely this $U(1)$ that is absent in
four dimensions, the resulting theory will again be non-chiral.  This
argument does not necessarily preclude the possibility of non-supersymmetric
chiral reductions.  However, that is not of present interest.

While the $M_4\times S^2$ reduction in itself is non-chiral, nevertheless
chirality may still survive in brane models \cite{quevedo2} 
where chiral families
live solely on the branes, and not in the bulk.  In fact, from a
braneworld perspective, the present model provides an alternative framework
to the Salam-Sezgin model, where here the bulk solution preserves ${\cal N}=2$
supersymmetry, and it is the branes themselves that provide both
chirality and an additional halving of supersymmetry to ${\cal N}=1$.
It would be of interest to study the resulting braneworld models constructed
from the present theory.

\section*{Acknowledgment}

     J.T.L. would like to thank the George P. \& Cynthia W. Mitchell
Institute for Fundamental Physics at Texas A\&M University for hospitality
during the course of this work.

\end{document}